\documentclass{emulateapj}
\usepackage{natbib}
\newcommand{\pacz}{Paczy\'nski}

\shorttitle{Candidate microlensing events from M31 observations
with the Loiano telescope}
\shortauthors{Calchi Novati et al.}

\begin{document}

\title{Candidate microlensing events from M31 observations
with the Loiano telescope}

\author{S.~Calchi Novati\altaffilmark{1,2},
V.~Bozza \altaffilmark{1,2},
F.~De Paolis\altaffilmark{3},
M.~Dominik\altaffilmark{4},
G.~Ingrosso\altaffilmark{3}, 
Ph.~Jetzer\altaffilmark{5},
L.~Mancini\altaffilmark{1,2}, 
A.~Nucita\altaffilmark{3},
G.~Scarpetta\altaffilmark{1,2},
M.~Sereno\altaffilmark{5}, 
F.~Strafella\altaffilmark{3},
A.~Gould\altaffilmark{6}\\
(The PLAN collaboration)\altaffilmark{7}}

\altaffiltext{1}{
Dipartimento di Fisica ``E. R. Caianiello'', 
Universit\`a di Salerno, Via S. Allende, 84081 Baronissi (SA), Italy}
\altaffiltext{2}{INFN, sezione di Napoli, Italy}
\altaffiltext{3}{Dipartimento di Fisica, Universit\`a del Salento and INFN, Sezione
di Lecce, CP 193, 73100 Lecce, Italy}
\altaffiltext{4}{Royal Society University Research Fellow,
SUPA, University of St Andrews, School of Physics \& Astronomy, North
Haugh, St Andrews, KY16 9SS, United Kingdom}
\altaffiltext{5}{Institute for Theoretical Physics,  University of Z\"urich, 
Winterthurerstrasse 190, 8057 Z\"urich, Switzerland}
\altaffiltext{6}{Department of Astronomy, Ohio State University, 140 West 18th Avenue, Columbus, OH 43210, US}
\altaffiltext{7}{http://plan.physics.unisa.it}

\begin{abstract}
Microlensing observations towards M31 are a powerful
tool for the study of the dark matter population in the form of MACHOs
both in the Galaxy and the M31 halos, a still unresolved issue, 
as well as for the analysis of the characteristics of the M31 luminous populations.
In this work we present the second year results of our pixel lensing campaign
carried out towards M31 using the $152~\mathrm{cm}$ Cassini telescope in Loiano.
We have established an automatic pipeline for the detection 
and the characterisation of microlensing variations.
We have carried out a complete simulation of the experiment
and evaluated the expected signal, including
an analysis of the efficiency of our pipeline.
As a result, we select 1-2 candidate microlensing events (according to
different selection criteria). This output is in agreement 
with the expected rate of M31 self-lensing events.
However, the statistics are still too low to draw definitive conclusions
on MACHO lensing.
\end{abstract}


\section{Introduction} 
The search for microlensing events aimed at the characterisation
of the MACHO distribution in galactic halos, first discussed by \cite{pacz86}, 
is by now an established technique. The results obtained up to now are,
however, debated. Towards the LMC the MACHO group
have claimed  the detection of a MACHO signal 
from objects of $\sim 0.4~\mathrm{M}_\odot$
that would constitute a halo mass fraction of 
about $f\sim 0.2$ \citep{macho00,bennett05},
whereas the EROS group have found no candidate microlensing events and
put a rather stringent \emph{upper} limit on the same quantity, 
$f<0.1$ in the MACHO mass range preferred
by the MACHO results \citep{eros07}. The issue of the nature of the detected
candidates still remains an open question \citep{sahu94,wu94,mancini04,novati06,evans06}.

The contradictory results obtained towards the Magellanic Clouds challenge
one to probe the MACHO distribution along different lines of sight.
Beyond the Galaxy, M31 represents the next most suitable target for 
microlensing searches \citep{crotts92,agape93,jetzer94}.
Looking at it from outside, we can globally study the M31 halo; 
the line of sight towards M31 allows one to probe the Galactic halo
along a different direction;
the inclination of the M31 disk is expected to give a clear signature
in the spatial distribution for microlensing events due to lenses in the M31 halo. 
Several observational campaigns have been carried out:
AGAPE \citep{agape97}, who presented
the first convincing microlensing candidate along this line of sight
\citep{agape99}, Columbia-VATT 
\citep{crotts96}, POINT-AGAPE \citep{point01,paulin03}, SLOTT-AGAPE 
\citep{novati02,novati03},
WeCAPP \citep{wecapp03}, MEGA \citep{mega04}, 
NainiTal \citep{nainital05}.
The detection of a few  microlensing candidates has
been reported, as well as first, 
though contradictory, conclusions on the MACHO content
along this line of sight. The POINT-AGAPE group
have reported evidence of a MACHO signal \citep{novati05}, whereas
the MEGA group have concluded that their detected signal is compatible
with the expected M31 self-lensing rate \citep{mega06}.
Very recently, \cite{arno08} have presented a new analysis
of a previoulsy reported bright event observed towards the M31 central region.
Taking into account the effects of the source's finite size,
they have concluded that the lens of this event 
should be attributed to the MACHO population.
Finally, we recall that a few interesting attempts have also been proposed \citep{totani03}, or
already carried out \citep{baltz04}, towards targets located beyond the Local Group.

In 2006 we began a new observational microlensing campaign towards M31
using the Cassini $152~\mathrm{cm}$ telescope  at the ``Osservatorio Astronomico 
di Bologna'' (OAB) located in 
Loiano\footnote{http://www.bo.astro.it/loiano/index.htm.}.
The results of the first year pilot season have been discussed in \cite{novati07}.
In this paper we discuss the second year campaign. As the main result,
we have carried out a complete analysis of the microlensing
flux variations, selected two  microlensing candidates and compared them
with the expected microlensing signal.
In Sect.~\ref{sec:data} we present the observational
setup and outline our data reduction and analysis technique.
In Sect.~\ref{sec:anaml} we present our pipeline for the search for microlensing-like
flux variations. In Sect.~\ref{sec:mls} we present the simulation of 
the experiment with an evaluation of the expected signal. In Sect.~\ref{sec:res}
we discuss the main results of the present analysis. Finally, in the Appendix
we describe in some detail some of the steps of our selection pipeline
and of our Monte Carlo scheme.

\section{Data analysis} \label{sec:data}

\subsection{Observational setup, data acquisition and reduction} \label{sec:setup}

The data have been collected at 
the $152~\mathrm{cm}$ Cassini Telescope located in Loiano (Bologna, Italy).
We make use of a CCD EEV of $1340\times 1300$ pixels of $0.58''$
for a total field of view of $13'~\times~12.6'$,
with gain of $1.0\mathrm{e}^-$/ADU (this value has changed
with respect to that  of the first season
because of some electronics problems) and low read-out noise (3.5 e$^-$/px).
We have been monitoring two fields of view around the inner  M31 region,
centered respectively in 
RA=$0{\mathrm h}42{\mathrm m}50{\mathrm s}$, 
DEC=$41^\circ 23' 57''$ (``North'')
and RA=$0{\mathrm h}42{\mathrm m}50{\mathrm s}$, 
DEC=$41^\circ 08' 23''$ (``South'') (J2000),
so to leave out the innermost ($\sim 3'$) M31 bulge region,
and with the CCD axes parallel to the south-north and east-west directions
so to get the maximum field overlap with previous campaigns.
This second year  campaign lasted 50 consecutive full nights,
from November 11 to December 31, 2007, with a fraction of good weather 
of almost 60\%.
In order to test for achromaticity, data have 
been acquired in two bandpasses (similar to Cousins $R$ and $I$),
with exposure times up to 6 minutes per frame.
Overall we collected about 410 (280) exposures per field
over 31 nights in the $R$ ($I$) 
band\footnote{We do not have 
exactly the same number of data points per night per filter
for the two fields, so that the indicated value
is actually an upper limit. Furthermore, within the analysis,
we exclude a small fraction of data points
that show anomalously large relative error values,
usually associated with poor seeing conditions or, more
generally, poor image quality.}.
Typical seeing values are $\sim 2''$
(somewhat worse than during the first season).
Sky flat frames were taken whenever possible
so as to build a master flat image (per filter),
and standard data reduction, including bias subtraction, 
was carried out using the  IRAF package\footnote{http://iraf.noao.edu/.}.
We corrected $I$ filter data for fringe effects.
The analysis presented in this paper is based on the 2007 season
data only.

\subsection{Image analysis} \label{sec:image}

As for the preliminary image analysis
we closely follow the strategy (the ``pixel-photometry'')
adopted by the AGAPE group
\citep{agape97,novati02}, wherein each image
is geometrically and photometrically aligned
relative to a reference image.
To account for seeing variations we then substitute
the flux of each pixel with that of the corresponding 
5-pixel square ``superpixel'' centered on it 
(whose size is chosen so to cover most of the average seeing disk)
and then apply an empirical, linear, correction in the flux,
again calibrating each image with respect to the reference image.
The final expression for the flux error accounts both
for the statistical error in the flux count
and for the residual error linked to the seeing correction procedure.
Finally, in order to increase the signal-to-noise ratio,
we combine the images so to get 1 data point per night per filter.

We evaluate the calibration zero point for the instrumental magnitude
versus standard $(RI)_C$ magnitudes by using a sample
of secondary reference stars \citep{massey06}. We find
for $R$ and $I$ bands data $C_R=23.1$ and $C_I=22.7$, respectively
(the reported values  corresponding to the standard magnitude for an object
with instrumental magnitude of $1~\mathrm{ADU~s}^{-1}$).

\section{Microlensing event search pipeline} \label{sec:anaml}

We have established a fully automated pipeline for the
detection and the characterisation of microlensing-like flux variations.
We work in the ``pixel-lensing''  regime \citep{gould96},
in which one looks for flux variations whose
sources are not resolved objects, so that one has to monitor flux variations
of every element of the image, further characterized by the fact
that the noise is dominated by the underlying background level
(the varying M31 surface brightness).
As for this specific analysis,
our strategy starts from that described in \cite{novati05}
with a few changes introduced to take into account the peculiarities
of the present data set.

During the analysis we have to face two main sources of contamination:
``fake'' signals, namely spurious variations to be attributed to cosmic rays,
defects in the CCD, saturated pixels and so on;
background intrinsically variable objects, that can either mimic microlensing
signals or, somewhat more dangerously, add non-Gaussian noise to the light curves.

Our pipeline can be schematically divided into four steps.
First: detection of the potentially interesting flux variations.
Second: characterisation of the light curve shape.
Third: probe against the contamination by spurious detections.
Fourth: probe against the contamination by variable signals.

\subsection{Bump detection} \label{sec:bump_detection}

As for the first step we closely follow the strategy outlined
in \cite{novati03,novati05}. To begin, we detect  flux variations 
along light curves using the $L$ estimator
(we ask $L>50$ to get rid of too small S/N variations). 
Each given flux variation enhances a signal over a few
pixels (a ``cluster'') around the central one. 
We make use of the ``$Q$'' estimator to characterize the significance
of the selected flux variations\footnote{It results $L>0$ whenever there
are at least three consecutive points at least $3\sigma$ above
the background. The value of $Q$ is given by the ratio
of the $\chi^2$ of a flat baseline fit over that of a \pacz\ fit \citep{novati03}.}. 
We  fix a lower threshold $Q_\mathrm{th}=50$. 
At this stage, therefore, we have to shift from
the light curve analysis (the estimation of $L$ and $Q$),
to an analysis based on the spatial information
across the CCD in which we have to distinguish, separate and pick up the 
flux variations associated to each different cluster.
This search  is somewhat biased in favour of light curves showing a single
variation (in particular, all short period
variables are in principle excluded).
This first step is carried out using the (more numerous and
less noisy) $R$ band data only.

\subsection{Light curve shape} \label{sec:light_curve_shape}

The aim of the second step is to single out the variations
whose shape is compatible with that of a \pacz\ light curve.
To this end, we use a series of selection criteria.
As a starting point we perform a 7-parameter
\pacz\ fit in the two bands 
simultaneously\footnote{We use the CERNLIB-MINUIT libraries, 
http://cernlib.web.cern.ch/cernlib/.}. As an output
we retain the following parameters: 
the baseline levels; the time of maximum magnification, $t_0$;
the full-width-at-half-maximum duration, $t_{1/2}$ 
(proportional to the Einstein time, $t_\mathrm{E}$, multiplied by a function
of the impact parameter $u_0$); 
the flux deviation at maximum (with respect to the baseline),
which we convert to magnitude and denote $\Delta R$, together with the color 
of the variation, $R-I$, ($\Delta R$ is proportional to
the source  flux, $\phi^*$, multiplied by a function of the impact parameter);
and the reduced $\chi^2$ of the fit. 
We make use of the full parametrisation of the \pacz\ fit,
looking for the Einstein time, the magnification and 
the unlensed source flux values,
even if the intrinsic parameter degeneracy (linked to the fact
that the source is not resolved),
in most cases, does not allow one to accurately
estimate the single parameters. 
As a selection criterion we ask $\chi^2/\mathrm{dof}<10$. This rather high threshold
is motivated by the necessity to handle 
light curves contaminated by low level noise of non-Gaussian nature
that can be attributed in particular to nearby blended intrinsic variables.

As a second test on the shape, we ask for the bump to be suitably sampled
by the observed data points. 
We split our analysis on the basis of a more or a less demanding requirement,
so that we are going to refer to set ``A'' and set ``B'' of candidates, respectively.
The details are given in the Appendix~\ref{sec:app_t0}.

As a final test on the shape, we look at
the characteristics of the detected flux variations.
The two relevant parameters are the event duration,
$t_{1/2}$, and the flux deviation at maximum, $\Delta R$.
In order to appropriately delimit these parameter spaces,
we have to balance for the efficiency,
the expected event characteristics and the risk of contamination
of the background of variable stars. 
We introduce a cut to exclude too faint variations, too noisy
and therefore difficult to distinguish against the 
variable contamination. As a selection criterion, we ask for $\Delta R<21.5$.
As for the duration, we do not expect microlensing events to last more than
10-20 days (Sect.~\ref{sec:model}). Besides, we expect
long-duration variations to be heavily contaminated by intrinsic variable signals. 
However, we prefer not to introduce any cut for this parameter allowing for long
duration candidates. As detailed below, all of these are rejected in
the following steps of the analysis anyway.

\subsection{PSF shape: spurious detection} \label{sec:psf_shape}

Up to now the analysis is based on the light curve pixel-photometry
only (besides the initial ``cluster'' analysis),
in particular we do not make use of the PSF of the flux variations 
we are looking for. In this respect our approach is completely
different from that based on the difference 
image analysis.
As it is, however, this approach suffers from a high risk
of contamination by spurious variations. To reject them
in an automated way, as a third step we perform an extremely rough
difference image analysis. The underlying rationale is that,
whenever we detect a variation on a light curve,
in order to retain it we want to ``see'' a corresponding well shaped PSF
when we look at the image difference of the maximum magnification minus the baseline.
Futher details are given in the Appendix~\ref{sec:app_psf}. 

\subsection{Variable signals} \label{sec:variable_signals}

As a fourth and final step we probe the surviving variations
against the background of variable contamination.
Our limited baseline, 50 days, does not allow us by itself
to carry out this programme. For this reason we make use
of the 3-years baseline of the 
POINT-AGAPE data set\footnote{Data collected at the 2.5m INT telescope
during 1999-2001 \citep{point01,paulin03}.}.
The rationale is that we want to reject flux variations
that show variability along  the much longer INT baseline
with a comparable flux deviation to that detected on our OAB data.
As a first step, given the OAB detected variation, 
we look for the corresponding pixel
within the POINT-AGAPE data set\footnote{The accuracy of the
astrometric trasformation is below 1 pixel, but we must accept the limit
given by the larger size of the OAB pixels, 0.58'', with respect to the
POINT-AGAPE ones, 0.33''.}. To probe variability along the INT light curve
we  use a Lomb periodogram analysis.
As an estimator, we use the power ``$P_R$'' as defined in Numerical Recipes \citep{numrec92}.
Second, we test whether the INT and the OAB flux deviations are compatible. 
To this end, given a relative  flux calibration 
between the two data sets, first we rescale the OAB flux deviation;
then we evaluate the difference
of the observed flux deviations (the rescaled OAB minus the INT one),
normalized by the INT error. We take this quantity, 
which we define $\delta(\Delta f)$, as our estimator.
As a selection criterion, we ask for $P_R<20$ \emph{or} $\delta(\Delta f)>5$.

\section{The expected microlensing signal} \label{sec:mls}

In order to gain insight into the observed signal,
first of all we have to evaluate the expected signal for our
experimental set up. 
We briefly outline our approach as follows \citep{novati05}. First, 
we run a Monte Carlo simulation whose
purpose is to give us the characteristics, and in particular the number,
of the expected microlensing events. To this end,
we have to specificy an astrophysical model and
a model for the microlensing magnification, besides
reproducing as closely as possible
the actual experimental conditions. As for this last point,
to account for those aspects of our pipeline that can not be accurately 
reproduced within the Monte Carlo, we carry out a simulation
on the real data set of the events selected within the Monte Carlo.

\subsection{The model} \label{sec:model}

We consider  the bulge and the disk populations of M31 as sources,
both M31 bulge and disk stars as lenses (we  refer to these
events as ``M31 self lensing''), and MACHOs in both the M31 and the Milky Way
dark matter halos. We assume an M31 distance of $770~\mathrm{kpc}$.

As a model for the M31 luminous components we take the \cite{kent89}
bulge-disk decomposition, including the bulge ellipticity,
and with the missing information of the vertical distribution of the disk 
modeled with a $\mathrm{sech}^2$ law and scale height of $300~\mathrm{pc}$.
For both  halos we assume a spherical isothermal distribution
with a core radius $a=5~\mathrm{kpc}$.

The peculiarity of M31 microlensing is that we look for flux variations
of unresolved sources. The issue of estimating
the number of available sources is therefore particularly delicate.
To this end, we consider the value of the M31 surface brightness 
(as given by \citealt{kent89}) and the underlying 
luminosity function. As has already been remarked 
\citep{agape97}, it turns out that only 
luminous sources (about $M_I<2$) are expected to give rise to detectable events.
In the most crowded region, this may sum up to hundreds of available
source stars \emph{per pixel} (this can be considered as a completely blended
situation with respect to, for instance, studies towards the LMC).
As for the luminosity function that we use to characterize the \emph{sources}, 
it is worth stressing that
we are most interested in its bright end (even if we need
the information over the complete magnitude range for the normalisation),
whereas, for the mass function that we need for the luminous \emph{lenses},
we are rather interested into the opposite tail.

We have made use of the IAC-star software \citep{iac04} to build a synthetic
luminosity function for the M31 bulge, following the procedure
outlined in \cite{bozza08}, in particular as for the metallicity
distribution \citep{sarajedini05}, 
with the difference that we have now used, as a mass function,
a power law $\xi(m)\propto m^{-\alpha}$ with index $\alpha=1.33$ 
up to $1~\mathrm{M}_\odot$ and $\alpha=2$ above. 
For bulge lenses we have for consistency used the same mass function
with upper bound fixed at one solar mass.
For the disk luminosity function, as in \cite{novati05},
we make use of the local neighborhood data obtained by Hipparcos
corrected at the bright end \citep{perryman97,jahreiss97}. For the disk lens
mass function we follow as well the local determination \citep{kroupa07}
with upper bound fixed at $10~\mathrm{M}_\odot$.
For MACHO masses we try a set of single values ranging from
$10^{-3}$ to $1~\mathrm{M}_\odot$.

We fix the total mass of the 
bulge to $4\times 10^{10}~\mathrm{M}_\odot$
\citep{kent89} (this can be considered a ``safe'' value for microlensing analyses,
for the purpose of evaluating the expected self-lensing signal,
as it is likely to be an upper limit \citep{arno06} for this quantity), 
and that of the disk to
$3\times 10^{10}~\mathrm{M}_\odot$ \citep{kerins01,arno06}. 
Looking for microlensing effects, the value we are actually
interested in is the \emph{stellar} mass, and indeed
our overall value for this quantity
agrees well with the analysis of \cite{tamm07a}. 
We consider a uniform extinction
across the field, both foreground, $E(B-V)=0.062$ \citep{schlegel98},
and intrinsic (for this second term, this hypothesis
should of course be taken only as a first order approximation), 
$\mathrm{ext}_R=0.19$ for the bulge \citep{han96,arno06}
and $E(B-V)=0.22$ for the disk \citep{stephens03,arno06}.
Together with the M31 color (not corrected for extinction) 
$B-r=1.3$ \citep{kent89}, $B-V=1.0$ and $V-R=0.8$ \citep{walterbos87},
this translates into the values (corrected for extinction) 
$M/L_R= 3.1,\,1.1$, for the bulge and disk respectively.

The bulge velocity distribution is dominated by its dispersive component,
with line of sight velocity dispersion of $\sigma=120~\mathrm{km~s}^{-1}$.
For the disk we take into account both the dispersive motion, 
$\sigma=60~\mathrm{km~s}^{-1}$ (a value that can be taken as un upper limit),
and a circular bulk motion, with disk circular velocity 
$v=250~\mathrm{km~s}^{-1}$ \citep{carignan06}.
M31 proper motion is set according to \cite{vdm07}.

Given the values of the circular velocity,
$220~\mathrm{km~s}^{-1}$ and $250~\mathrm{km~s}^{-1}$
for the Milky Way and M31 respectively,
we fix accordingly the central densities and therefore
the overall masses of the halos 
and the values of the one dimensional velocity dispersions.
As for the total dark matter halo mass value, within 
a truncation radius of $100~\mathrm{kpc}$ and $130~\mathrm{kpc}$
for the Milky Way and M31 (we estimate the ratio of these values 
from those of the circular velocities),
we have respectively $1.0\times 10^{12}~\mathrm{M}_\odot$ 
(in good agreement with previous determination, e. g. \cite{vallenari06},
but somewhat in excess with respect to the recent determination by \citealt{xue08})
and $1.8\times 10^{12}~\mathrm{M}_\odot$.

\subsection{The simulation} \label{sec:efficiency}

\begin{figure}
\epsscale{.90}
\plotone{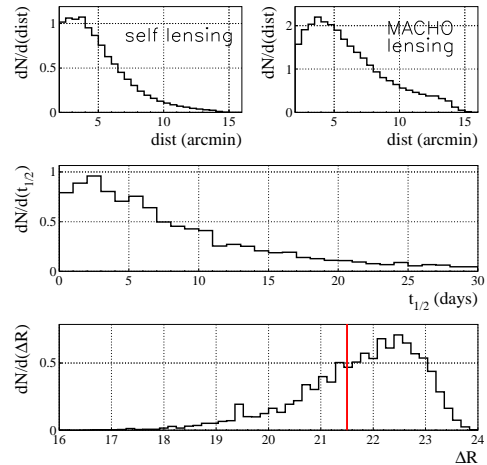}
\caption{
Characteristics of the expected events from the Monte Carlo analysis.
From top to bottom: histograms for the distance from the M31 center,
duration (full-width-at-half-maximum) 
and the flux deviation at maximum magnification. 
Histograms for event duration and flux deviation are given
for self lensing only.
For all of the histograms we show all of the events selected,
namely for the full range of the parameters.
In  the bottom panel, the solid line
indicates the threshold value we use in the pipeline.
\label{fig:mls}
}
\end{figure}

Within the Monte Carlo simulation, given the astrophysical model, we 
generate microlensing signals with \pacz\ magnification
corrected for finite size effects of the 
sources\footnote{For bulge sources we use
the radius values from the IAC database,
for disk sources we use a color temperature
relation evaluated from the model
of \cite{robin03} and we evaluate the radii from
Stefan's law using a table of bolometric corrections
from \cite{murdin01}.  For the microlensing amplification
we use the analytical expression derived in \cite{witt_mao_94}.},
build the correspoding light curves and carry out
a first rough selection,  paying attention
not to reject any light curve that could be detected
through our pipeline. In particular, for a variation to be selected,
as a unique criterion we ask for at least 3 consecutive points
(with one point per night and reproducing the observed sampling)  
3 sigma above the backrgound level
($L>0$ according to the estimator introduced in Sect.~\ref{sec:anaml}).
On the other hand, we are aware that 
we can not reproduce within the Monte Carlo,
where we only deal with \emph{light curves}, 
all of the the actual conditions of the pipeline we 
carry out in the real data set (where the analysis
starts from the \emph{images}).
Amongst other effects, we most prominently
can not reproduce crowding effects, the underlying variable signals and
the sources of non-Gaussian noise.
Furthermore, we can not run the first essential step
of bump ``cluster'' detection.
To account for these effects  we simulate those light curves
that are selected within the Monte Carlo on the images
(\emph{before} the geometrical and all 
photometric corrections, namely, on the astronomical images
just after the basic bias-flat fielding reductions) and then we run
from scratch our pipeline. Conceptually, this is just
a last step in the Monte Carlo that allows us to accurately evaluate
the efficiency of our pipeline. The ``efficiency'', hereafter, should therefore
be intended as that relative to the light curves selected within the Monte Carlo. 

Within the Monte Carlo each simulated event carries a (different) ``weight'',
$w_i$ (where $i$ is the index of the simulated events)
that is linked in part to the drawing process and in part 
to the quantity we are evaluating (the microlensing rate),
and is completely independent of the selection 
process\footnote{As for this technical aspect, our analysis
therefore differs, for instance, from that discussed in \cite{kerins01},
as we draw all the values of the random variable according to their actual
distributions rather than according to the microlensing rate
(further details are given in Appendix~\ref{sec:app_mls}).
In addition, \cite{kerins01} propose to perform the Monte Carlo simulation
only to evaluate the pipeline efficiency whereas we make use
of this tool also to evaluate the number of expected events.}.
The expected number of events
is therefore the sum of the weights, $n_\mathrm{exp} = \sum_i w_i$, where
the sum runs over the events selected within the Monte Carlo
(correspondingly we can estimate the associated statistical error
based on Poisson statistics).
Accordingly, by ``efficiency'', $\epsilon_w$, we mean
the ratio of selected over simulated events, where
the number we refer to is always given by the sum of the weights.
This is usually different (both numerically and from a logical point of view)
from the actual ratio of selected over simulated events,
a quantity that is not used even if
in some cases it may be useful to be looked at.
In particular, we do not expect, and in fact we do not find, 
these two values to be too different. Indeed, such a result  should
be taken as a hint of the presence of some bias
in the way the event weights are distributed 
with respect to the selection process within the simulation.

For each lens population, we simulate up to a few thousands events per field,
with 500 events per field per simulation in order to avoid overlap problems.
Indeed, in particular in the inner M31 region, where we expect most of the events,
simulated events may overlap (we draw at random from the Monte Carlo
the events we simulate, with all their characteristics, including
the line of sight) and thus lead us to bias the estimate of the 
efficiency of our pipeline.
To test for this effect, for a fixed set of 500 events
per field for which we had already evaluated the efficiency,
we have carried out as many different simulations as needed, 
taking care to leave a minimum distance of at least 20 pixels among
any couple of generated events, so to exclude overlap problems.
As a result, we have found no significant differences in the two analyses.
Overall, we have simulated 12000 light curves to evaluate self-lensing 
efficiency, and 8000 for each value of the mass for MACHO lensing.

According to the selection pipeline, in which we require
for the flux variations to be large enough, $\Delta R<21.5$,
out of the Monte Carlo 
we extract, and then simulate 
on the images, selected events with flux deviation at maximum 
down to\footnote{As a matter of fact, within
the Monte Carlo, where we have only a statistical error,
we find a much fainter ``theoretical'' lower bound for the flux deviation
at maximum (Fig.~\ref{fig:mls}). The efficiency analysis, on the other
hand, showed us that the choosen treshold value
for $\Delta R$ is appropriate, because the efficiency dramatically
decreases when we consider too faint flux variations.}
$\Delta R=21.8$. This limit 
is used accordingly when we evaluate the number of expected events.
This way we allow for the observed rms of the evaluated flux deviations
versus the input values. The exact value of this threshold
is not, however, essential, as long as we keep
the selection criterion on the flux deviation 
fixed at $\Delta R<21.5$ coherently with the selection pipeline.
A brighter threshold would translate into a larger  
value for the efficiency and, at the same time, 
a smaller number of expected events (not corrected for the
efficiency), and vice versa. These two effects balance
when we evaluate the number of expected events corrected for the efficiency.

As for the expected characteristics of the observed events,
in Fig.~\ref{fig:mls} we show the resulting distributions
for the distance from the M31 center (both self lensing and MACHO lensing),
and, for self lensing, the distribution for 
the durations (the full width at half maximum $t_{1/2}$)
and the flux deviations at maximum ($\Delta R$).

Finally, it is worth stressing that we 
simulate microlensing events only.
Therefore the simulation is restricted
to saying whether and how our pipeline is going to select 
microlensing signals but it can say nothing about whether 
it might select as a microlensing
a variable signal of different origin.

\section{Results} \label{sec:res}

In this Section we present and discuss the results of our analyses: the selection
pipeline for microlensing flux variations and the evaluation 
of the expected microlensing signal.

\subsection{The  selection pipeline} 

\begin{figure}
\epsscale{.90}
\plotone{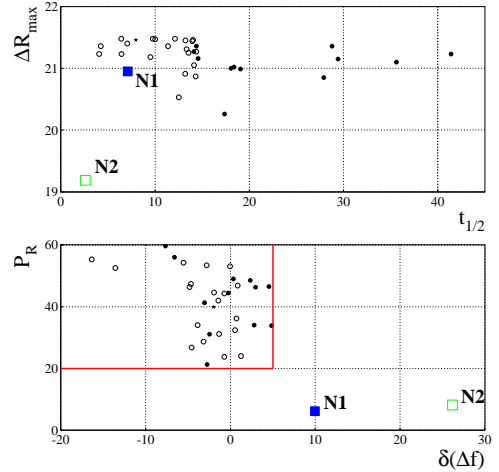}
\caption{
Selection pipeline results: 
the sample shown here consists of the light curves
selected after the PSF analysis.
Scatter plot of the flux deviations  versus the event duration (top panel);
scatter plot of the Lomb ``power'' $P_R$
versus the relative flux difference $\delta(\Delta f)$
(filled and empty symbols for set A and B variations,
respectively). The squares indicate the two candidate
microlensing events of our selection (according
to the terminology introduced in Sect.~\ref{sec:gold}).
The solid lines delimit the excluded region (bottom
panel, upper left corner) in the last step of the analsyis.
The star symbol indicates the second event, beside N1,
 belonging to both set A and set B.
Bottom panel : for visualisation purposes we do not
show a few rejected events with $\delta(\Delta f)<0$
and $P_R>20$.
\label{fig:selection}
}
\end{figure}

\begin{figure}
\epsscale{.90}
\plotone{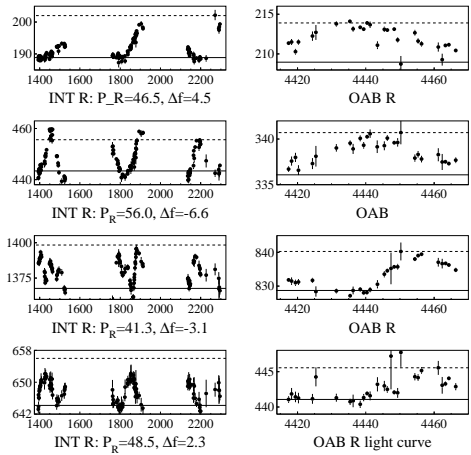}
\caption{
INT and OAB $R$-band light curves extension of 4 out of the 14 set A 
flux-variations OAB candidate microlensing events selected after the shape
and the PSF analyses (left and right panels, respectively). The solid
lines mark the evaluated baseline level, while the dotted lines show
the maximum flux deviation value along the detected OAB flux variation (right panels)
and this same value once rescaled on the INT data set (left panels). 
For each INT data light curve we indicate
the result of the Lomb periodogram analysis, the power $P_R$,
and the difference of the flux deviation along the
OAB and the INT light curves, $\delta(\Delta f)$.
The units on the $x$ axes are time in days (JD-2450000.0).
The ordinate axes units are flux in ADU s$^{-1}$ per superpixel. 
\label{fig:var}
}
\end{figure}

\begin{deluxetable}{rrrrr}
\tablecolumns{5}
\tablewidth{0pc}
\tablecaption{Microlensing selection pipeline : the results\label{tab:pipeline}}
\tablehead{
\colhead{} &\multicolumn{2}{c}{selection} & \multicolumn{2}{c}{simulation}\\
\colhead{criterion} & \multicolumn{2}{c}{\# events} & 
\multicolumn{2}{c}{efficiency $\epsilon_w$ (\%)}\\
\colhead{} & set A & set B & set A & set B\\
}
\startdata
bump detection &\multicolumn{2}{c}{4200}& \multicolumn{2}{c}{$41.2\pm3.5$} \\
$L_1>50$ &\multicolumn{2}{c}{3033}& \multicolumn{2}{c}{$35.5\pm3.2$} \\
\hline
shape analysis&&&\\
$\chi^2/\mathrm{dof}<10$ &\multicolumn{2}{c}{2901} & \multicolumn{2}{c}{$32.7 \pm 3.0$}\\
sampling & 174 & 241 &  $18.3 \pm 2.0$ & $21.9 \pm 2.4$\\
$\Delta R<21.5$ & 75 & 108 &  $13.0 \pm 1.5$ & $20.4 \pm 2.3$\\
\hline
PSF & 14 & 23 & $13.0 \pm 1.5$ & $15.7 \pm 1.9$\\
\hline
variable & 1 & 2 & $11.9 \pm 1.4$ & $14.6 \pm 1.8$
\enddata
\tablecomments{The results of the selection pipeline for microlensing
light curves: analysis and simulation.
For each step we report the number of selected light curves and
the efficiency of the pipeline (\%) for the expected self-lensing signal.
According to the choice of the sampling criterion,
we have split our selection results in set A and set B (left and right column,
respectively).}
\end{deluxetable}

In Table~\ref{tab:pipeline} we report the results of the selection pipeline analysis
together with the results for the efficiency
of the corresponding analysis carried out on self-lensing simulated events.

We start the analysis working over the complete set of pixels,
namely $2\times (1340\times 1300)$ light curves.
The initial sample of selected flux variations consists
of $\sim 4000$ light curves. Within the shape analysis,  the sampling cut severely
reduces this initial set, and then the  flux deviation cut
leaves us with $\sim 200$ light curves, most of which,
according to the PSF analysis, are to be attributed to spurious variations.
Finally we are left with $\sim 40$ flux variations,
divided into sets A and B (according to the sampling criterion),
most of which we expect to be intrinsic variables
whose single-bump appearance is to be attributed to
our short (50 days) baseline. 
(As outlined in Appendix~\ref{sec:app_t0},
set A flux variations are not a subsample
of set B: among those surviving the PSF cut, 14 and 23 respectively,
only two flux variations are in common between the two data sets,
out of which one also survives the last cut.) 
The results of the last cut analysis 
are shown in the bottom panel of Fig.~\ref{fig:selection}
in the parameter space $P_R$-$\delta(\Delta f)$.
We find, as expected,
the flux deviation of most OAB variations to be compatible with
the corresponding INT variations (small values of $\delta(\Delta f)$),
for which, at the same time, we find a clear sign of variability (large value
of $P_R$). As an example, in Fig.~\ref{fig:var}, we show the INT extension
of 4 OAB selected light curves. Only for two
selected OAB flux variations, instead, we find the corresponding
INT extension to be flat. Therefore the selection pipeline
finally leave us with two microlensing candidates
(one belonging to set B only).
The same sample of light curves is represented
in the $\Delta R_\mathrm{max}-t_{1/2}$ parameter space
(top panel of Fig.~\ref{fig:selection}).
We find most of the variations located in the upper part 
(corresponding to small flux deviations),
with set B variations (empty symbols) biased towards short durations.
In particular we find the set A microlensing candidate 
(filled square symbol) located in a parameter space region 
where the contamination by the intrinsic variable signals is large.
The only clear outlier is the
bright and short set B microlensing candidate.
We discuss the selected candidates in detail 
in Sect.~\ref{sec:gold}. 

Comparing with the efficiency simulation analysis,
a few points are worth being mentioned.
First, it may look as if the ``bump detection'' 
step alone severely reduces the overall efficiency.
However, in fact, only a very small fraction
of light curves that are not selected at this point
would pass all the other criteria. 
According to the same principle, the single cut
that excludes most of the simulated light curves
is the sampling criterion\footnote{The overall
efficiency would jump to $\sim 20\%$ taking into account all the criteria
except the sampling criterion.}. This is also the reason
why we have split our selection pipeline
at this level. As for the simulation, we stress that
this simply reflects the choice we have made in the Monte Carlo
to select  light curves on the basis of the $L>0$
criterion only. Next, the PSF analysis proves
to be a rather efficient criterion.
Indeed it results that almost 50\% of  the 
simulated light curves fulfill this criterion
and that this fraction rises to about 80\% if we consider
the subset of light curves that have already passed
the bump detection and the shape analysis cuts.
A usual reason that may cause the PSF Gaussian fit to fail
is the presence, near the simulated event, of some other
resolved object. Often enough, however, in these cases
also the pixel photometry we use may have problems.
(For both of these related aspects, a proper difference
image analysis approach would be of course expected to give
better results). Poorly sampled light curves, on the other hand, 
simply do not have enough points near maximum
magnification, so that it turns out to be usually not possible to carry out
a good enough PSF fit. 
Finally, as for the analysis on the POINT-AGAPE extension
to check for variable signals, we have seen this cut to be essential
to get rid of otherwise dangerous contaminating flux variations.
At the same time, this shows to be an extremely efficient criterion.
Out of  the complete set of simulated light curves, only about 13\% of the INT
light curves are clearly variable ($P_R>20$)
and this fraction falls to 7\% when we add the demand 
for the INT flux deviation to be compatible with the
OAB simulated one.

\begin{deluxetable}{rrr}
\tablecolumns{3}
\tablewidth{0pc}
\tablecaption{Efficiency analysis: the results for MACHO lenses.\label{tab:eff}}
\tablehead{
\colhead{mass ($\mathrm{M}_\odot$)} &\multicolumn{2}{c}{efficiency $\epsilon_w$ (\%)}\\
\colhead{} & \colhead{set A} & \colhead{set B}\\ 
}
\startdata
1   & $16.5\pm 1.2$ & $18.7\pm 0.9$\\  
0.5 & $18.5\pm 1.1$ & $21.3\pm 1.2$\\  
0.1 & $16.5\pm 1.2$ & $19.9\pm 1.4$\\  
$10^{-2}$ & $13.8\pm 1.2$ & $16.5 \pm 1.3$\\  
$10^{-3}$ & $11.1\pm 1.3$ & $13.2 \pm 1.4$\\  
\enddata
\end{deluxetable}

In Table~\ref{tab:eff}, we report the results for the
efficiency of the simulations for MACHO lenses.
With respect to self-lensing events there is a (rather small)
effect linked to the different spatial distribution.
The main effect is, however, due to the value of the mass,
which is linked to the event duration, with smaller
values of the efficiency corresponding to decreasing
values of the MACHO lens mass.

In Tables~\ref{tab:pipeline} and \ref{tab:eff}
we have given the results for the efficiency 
as a single value for the overall set of simulated events.
On the other hand, the efficiency does vary quite significantly
for data binned, for instance, in the distance from the M31 center
and/or in the flux deviation at maximum. We find
the larger values for the efficiency, up to 30\% or more,
for bright events in the outer regions of M31. On the other hand,
the expected number of events, not corrected
for the efficiency, is larger near the M31 center
for faint events (Fig.~\ref{fig:mls}), namely, right where
the efficiency is smaller (down to below 5\%, depending
on the choice of the binning). Overall, however, we find
the expected number of events corrected for the efficiency
to be rather insensitive, within the statistical error
of the simulation, to any binning scheme, and this
motivates our choice for the way to present our results.

\subsection{The microlensing candidate events} \label{sec:gold}

\begin{figure}
\epsscale{.90}
\plotone{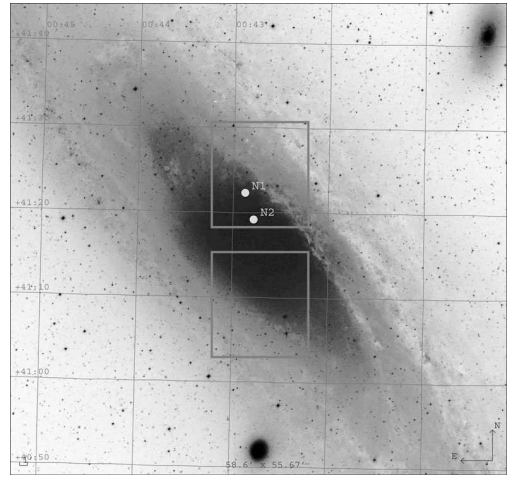}
\caption{
Projected on M31, 
we display the boundaries of the two 13'x12.6'
monitored fields. The filled circle mark the
position of the selected microlensing candidate events. 
\label{fig:m31}
}
\end{figure}

\begin{deluxetable}{rrrr}
\tablecolumns{4}
\tablewidth{0pc}
\tablecaption{Characteristics of the microlensing
candidates OAB-N1 and OAB-N2.\label{tab:gold}}
\tablehead{
\colhead{} & \colhead{OAB-N1} & \multicolumn{2}{c}{OAB-N2}\\ 
\colhead{} & \colhead{} & \colhead{OAB data} & \colhead{source flux fixed}\\
}
\startdata
$\alpha$ (J2000) & 0h 42m 57s &  \multicolumn{2}{c}{0h 42m 50s}\\
$\delta$ (J2000) & $41^\circ 22' 50''$ & \multicolumn{2}{c}{$41^\circ 18' 40''$}\\
$d_{\mathrm{M}31}$ (arcmin) & 7.1 & \multicolumn{2}{c}{2.8}\\
\hline
$t_0$ (JD-2450000.0) & $4433.8\pm 0.2$ & $4467.3_{-1.3}$ & $4466.6\pm 0.3$\\
$t_{1/2}$ (days) & $7.1^{+1.4}_{-1.3}$ & $2.6^{+1.0}_{-0.5}$ & $2.56^{+0.12}_{-0.20}$\\
$\Delta R$ & $21.1\pm 0.2$ & $19.1^{+0.8}$ & $19.5\pm 0.2$\\
$R-I$ & $1.0\pm 0.2$& $1.1\pm 0.1$ & 1.2\\
$\chi^2/\mathrm{dof}$ & 3.9 & 1.4 & 1.1\\
\hline
$t_\mathrm{E}$ (days) && $3.9^{+2.7}$ & $4.18^{+0.55}_{-0.41}$\\
$u_0$ && $0.23_{-0.12}$ & $0.224^{+0.008}_{-0.029}$\\
$\phi^*_R$ (ADU s$^{-1}$) && $15_{-12}$ & $7.6$\\
$\phi^*_I$ (ADU s$^{-1}$) && $28_{-22}$ & $15.1$\\
\hline
$R^*$ && $20.2^{+1.4}$ & $20.9$\\
$I^*$ && $19.1^{+1.7}$ & $19.7$
\enddata
\tablecomments{
For OAB-N2,
we show the results for the fit
performed both with our data set alone
(left column) and fixing the source flux
from a possible identification
with a source in the \cite{massey06} catalogue.
The error on the standard magnitude values and colors includes an extra $\pm 0.1~\mathrm{mag}$
term from the calibration equation. 
}
\end{deluxetable}

The selection pipeline described in the previous section
leaves us with two candidate microlensing
events, which we name OAB-N1 and OAB-N2 (``N'' indicates that
they are both located in our ``North'' field). 
Their characteristics are summarised in Table~\ref{tab:gold},
and their position within our field of view
is shown in Fig.~\ref{fig:m31}\footnote{The original 
M31 image has been taken from the CDS data base, http://cdsweb.u-strasbg.fr/.}.
As for the sampling criterion, OAB-N1 fulfills both
sets A  and B demands, while OAB-N2 only the set B one.

\begin{figure}
\epsscale{.90}
\plotone{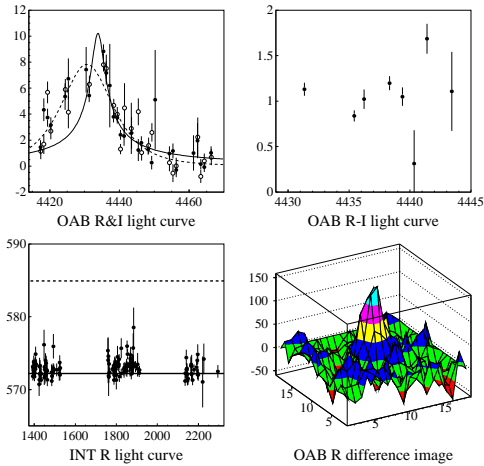}
\caption{
The microlensing candidate event OAB-N1. 
Top left panel: $R$ and $I$-band light curves
(filled and empty symbols, respectively).
The $R$-band data are normalized
with respect to $I$-band data using the color of the flux variation.
The solid and the dashed lines are the best-fitting \pacz\ curves
with $t_0=4433.8$ and $t_0=4430.6$ (JD-2450000.0), respectively
(Sect.~\ref{sec:gold} for details).
Top right panel:  color light curve.
Bottom panel (left): the 
extension along the INT data (the solid and dotted
lines are as in Fig.~\ref{fig:var}).
The units on the $x$ axes are time in days (JD-2450000.0).
The ordinate axes units are flux in ADU s$^{-1}$ per superpixel
(left panels) and magnitude (top right panel). 
Bottom panel right: OAB $R$-band data 
surface plot image difference between the image at maximum
magnification and the baseline level. The units on the
space $x-y$ axes is pixels, the surface plot values
are in ADU per pixel.
\label{fig:oab1}
}
\end{figure}

OAB-N1 : This is a relatively large flux variation, 
with significance bump estimators equal to $L=183$ and $Q=108$;
quite short, $t_{1/2}=7.1~\mathrm{days}$, and
not too bright, $\Delta R=21.1$. 
The OAB-N1 light curve is shown in Fig.~\ref{fig:oab1}, together
with its INT extension and the image difference 
around the candidate position upon which is based
our PSF analysis.
The INT data extension of the OAB-N1 light curve
appears to be flat (small Lomb periodogram power, $P_R=6.3$, and 
significantly large value for the difference 
of flux deviation, $\delta(\Delta f)=10.0$).
However, the quality of the \pacz\ fit is not good.
A few points before the bump deviate significantly 
from the expected shape, and this is reflected in the rather poor value
of  $\chi^2$. This might be attributed to 
underlying nearby variables, but we can not exclude 
it to be a sign of an intrinsic non-microlensing nature. 
Indeed, we have shown that OAB-N1 is located
in a part of the $t_{1/2}-\Delta R$ parameter space
where the background of variable stars is large
(top panel, Fig.~\ref{fig:selection}).
A possible contamination, compatible with
its ``one bump'' nature, besides a possible
very long period variable, might come from some kind of eruptive
variable (even if  Nov\ae~can be excluded 
because the flux variation is far too small).
On the other hand, the descent is fairly well sampled
and matches nicely enough the fitted \pacz\ shape.
In the top right part of Fig.~\ref{fig:oab1}
we show the ``color'' light curve, namely the ratio $R$ over $I$ band
of the difference of the light curve flux, along the bump,
and the background level, that we expect to be constant 
for  microlensing. 

Through the pipeline, and in particular for OAB-N1,
as an initial condition for the time of maximum magnification
in the \pacz\ fit we choose the value of
the time corresponding to the data point with the maximum
flux value, in this case $t=4435.4$ (JD-2450000.0).
During the fit procedure
it is difficult for this parameter to exit from
the $\chi^2$ minimum well around this value
(whose bounds are usually set by the sampling)
and indeed as a result we find $t_0=4433.8\pm 0.2$ (JD-2450000.0).
Motivated by the OAB-N1 light curve appearance,
irregular sampling and noisy aspect,
we have therefore carried out a search for other
$\chi^2$ minima in different regions of the $t_0$ space.
As a result we have found a new, well isolated, minimum
with a \emph{lower} value of the reduced $\chi^2$
($\chi^2/\mathrm{dof}=3.5$ to be compared with $\chi^2/\mathrm{dof}=3.9$
found in the previous case) for $t_0=4430.6\pm 0.4$ (JD-2450000.0).
Correspondingly, we find new values for the duration
and the flux deviation at maximum, $t_{1/2}=18~\mathrm{days}$
and $\Delta R=21.5$ (this last value being at the limit
of our threshold cut), namely, a rather different
result from the previous one. Such a light curve would have still been
selected as a microlensing candidate within our pipeline. However, the longer 
duration would have futher weakened its microlensing interpretation
(Sect.~\ref{sec:discussion}). 
The co-existence of these two minima 
might be suggestive, accepting the microlensing
hypothesis, of a binary lens or binary source solution.
The available data, however, do not allow us
to  robustly test this hypothesis\footnote{Through the selection pipeline 
we have only considered (single bump)
\pacz\ like microlensing variations. 
The analysis of the possible binary lens
solutions for OAB-N1, together with a 
systematic search for binary-like
flux variations, will be presented in \cite{bozza09}.}.

\begin{figure}
\epsscale{.90}
\plotone{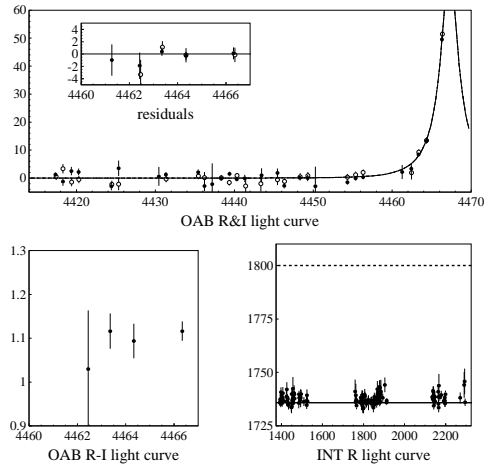}
\caption{
The microlensing candidate event OAB-N2.
Top panel: normalized $R$ and $I$-band light curves
(filled and empty symbols, respectively).
The $R$-band data are normalized
with respect to $I$-band data using the color of the flux variation.
The solid line is the best-fitting \pacz\ curve.
In the insert we show the residual with respect to the 
\pacz\ fit along the bump.
Bottom panels, left: color light curve;
right: extension along the INT data (the solid and dotted
lines are as in Fig.~\ref{fig:var}).
The units on the $x$ axes are time in days (JD-2450000.0).
The ordinate axes units are magnitude for the bottom
left panel, and flux in ADU s$^{-1}$ per superpixel for the remaining panels. 
\label{fig:oab2}
}
\end{figure}

OAB-N2 : This looks (Fig.~\ref{fig:oab2}) like an extremely bright and short
flux variation ($t_{1/2}=2.6~\mathrm{days}$,
$\Delta R = 19.1$), located near the M31 center
(at a distance $d=2.8'$), with a completely flat
INT extension ($P_R=8$, $\delta(\Delta f)=25$).
The sampling along the bump is, however, extremely poor
(indeed, both the time of maximum amplification and
the flux deviation at maximum are not strongly
constrained, Table~\ref{tab:gold}).
In particular, the lack of data points in the descent prevents us
from probing the expected microlensing symmetric shape,
so that it is difficult to draw definitive conclusions
about its nature. On the other hand, the rising part of OAB-N2 looks achromatic, 
and  is rather well constrained, and this strengthens
the microlensing hypothesis.
Indeed, both the shape and the achromaticity 
are hardly compatible with the only possible kind of contamination for
such a kind of flux variation, if not microlensing, 
namely some sort of eruptive-like object.
Furthermore, though few, the points along the bump 
allow us to get to a reasonable fit for all the parameters
of the microlensing event. At the $1~\sigma$ level, the $\chi^2$ analysis
gives us best values and \emph{lower} bounds for the impact parameter and the source flux
and  best value and \emph{upper} bound for the Einstein time (with, however,
a rather large relative $1~\sigma$ error, even exceeding 50\%, Table~\ref{tab:gold}).
Finally, the guess for the flux source value allows us to carry out a 
more constrained color analysis 
(as for OAB-N1, but subtracting the source flux from the baseline,
bottom left panel of Fig.~\ref{fig:oab2}).
As for the value of the source flux,
the results of the \pacz\ fit have been confirmed by the cross-identification
of the possible source star in the catalogue of \cite{massey06}.
Indeed, within 1 pixel of our evaluated position, we find
a typical red giant with $R_C=20.9$ and $R-I=1.2$
(values fully compatible with those evaluated from our data set alone).
The knowledge of the source flux allows us to better constrain the fit
parameters, in particular the time of maximum amplification
and the flux deviation at maximum (in fact $t_0$ shifts back
to the position of the last observed data-point so that,
accordingly, $\Delta R$ gets fainter). Furthermore, we may now completely
break the degeneracy among the amplification parameters. 
Indeed, we estimate $t_\mathrm{E} = 4.2^{+0.6}_{-0.4}$ days
and $u_0=0.22^{+0.01}_{-0.03}$ ($A_\mathrm{max}=4.5$).

We have also searched for $X$-ray counterparts in the \emph{XMM}-Newton 
archive\footnote{http://xmm.esac.esa.int/xsa/.}
and verified that 
neither of the microlensing candidates' coordinates
correspond to any of the identified sources in the EPIC images \citep{xmm01a,xmm01b,xmm01c}. 
In the case of OAB-N2 there is an $X$-ray counterpart, but only within
$\sim 30''$, so that we can safely rule out the identification.

\subsection{The expected number of events}

\begin{deluxetable}{rrr}
\tablecolumns{3}
\tablewidth{0pc}
\tablecaption{The expected number of microlensing events.
\label{tab:nexp}}
\tablehead{
\colhead{} & \multicolumn{2}{c}{$n_\mathrm{exp}$}\\ 
\colhead{} & \colhead{set A} & \colhead{set B}\\
}
\startdata
bulge-bulge & $0.35\pm 0.07$ & $0.44\pm 0.08$\\
bulge-disk & $0.18\pm 0.02$ & $0.23\pm 0.03$\\
disk-bulge  & $0.06\pm 0.01$ & $0.07\pm 0.01$\\
disk-disk & $0.015\pm 0.002$ & $0.019\pm 0.002$\\
\hline
SL & $0.6\pm 0.1$ & $0.8\pm 0.1$\\
\hline
mass ($\mathrm{M}_\odot$) &&\\
1 & $0.7\pm 0.1$ & $0.9\pm 0.2$\\
0.5 & $1.1\pm 0.2$  & $1.3\pm 0.2$\\
0.1 & $1.6\pm 0.1$ & $1.9\pm 0.1$\\
$10^{-2}$ & $2.0\pm 0.2$ & $2.3\pm 0.2$\\
$10^{-3}$ & $1.3\pm 0.2$ & $1.5\pm 0.2$
\enddata
\tablecomments{
The expected number of microlensing events, for M31 self lensing
(lenses belonging to either the M31 bulge or disk)
and MACHO lensing (lenses belonging to either the M31 or the
Milky Way halo), for a full halo with different values of the MACHO mass.
For self-lensing events we report also the number for
each lens-source population we consider.
We report the results of the Monte Carlo simulation
corrected for the efficiency evaluated
according to both set A and B criteria.
}
\end{deluxetable}

The Monte Carlo simulation, completed by the efficiency analysis,
allows us to estimate the expected number of microlensing events
for our experimental setup. In Table~\ref{tab:nexp} we report
the resulting values, already corrected for the efficiency.
The results we obtain using the set B criteria
is larger by up to about 30\% with respect to set A.
This is a combined effect of the larger value
of the efficiency and of the effective baseline length increase.

As for self-lensing events, it turns out that most of them,
$\sim 50\%$, are due to bulge-bulge events (lens-source, respectively), 
with the remaining distributed almost equally between 
disk-bulge and bulge-disk events. The second configuration is 
enhanced in the South field because in this case
we see the bulge in front of the disk. It is also
worth noting that, according to our model, almost half
of the overall bulge mass is located within our cone of view,
but only about 1/7 of that of the M31 disk.

As for MACHO lensing, about 2/3 of the events are expected to belong
to the M31 halo. Actually, the overall mass
within our cone of view of the M31 halo is larger than the 
Galactic one by a factor of a few thousands, 
but Milky Way halo lensing is strongly enhanced by the 
much larger size of the Einstein radius.
We expect the largest number of MACHO lensing 
events from $10^{-1}-10^{-2}~\mathrm{M}_\odot$
objects, as our efficiency 
drammatically drops for smaller MACHO masses.

Together with the expected event number we report
the statistical error associated with the simulation
on the basis of Poisson statistics. 
The uncertainties intrinsic
in the model (whose detailed discussion is beyond
the scope of the present paper) make, however, the corresponding
systematic error in principle even larger.

\subsection{Discussion} \label{sec:discussion}

We have designed our (fully automated) pipeline so as to get rid 
of most forms of stellar variability, 
while preserving genuine microlensing. As a result, therefore,
all survivors could in principle
be variables but with varying (hopefully large) degrees of confidence
that they are not.
Coming to the present data set, we have selected two flux variations
compatible with a microlensing signal.
While discussing the possible physical meaning
of this result, however, we have to keep in mind
that, although with different degrees of confidence,
we have no strong evidence in favour of
the microlensing hypothesis for either candidate:
OAB-N1 because of both the light curve appearance
(reflected in the $\chi^2$ value) and its position in the
duration/flux-deviation parameter space, and
OAB-N2 because of its extremely poor sampling
(that has not prevented us, however, from finding a convincing microlensing
\pacz\ fit further confirmed  by the possible identification
of the source). In conclusion, looking at the candidate light curves,
we might be tempted to classify OAB-N1 as a very ``poor'' candidate,
and OAB-N2 as a ``good'' one, keeping in mind, however, 
that this might simply reflect the bias induced by the observational sampling. 

With the care suggested by the above discussion,
we come now to the comparison of the observed events
with the expected signal. As for the event characteristics
we may take as a reference the distributions shown
in  Fig.~\ref{fig:mls}. Given the caveat that these distributions
can not be compared directly with the results of the analysis
because of the correlation
of the pipeline efficiency with the event characteristics,
both candidates appear compatible with the expected signal
with respect to  the duration and the distance from the M31 center
(being both shorter and nearer to the M31 center, OAB-N2 appears,
also in this respect, a stronger candidate than OAB-N1). 
Coming to the number of events, we must compare 
our observed candidates with
the results reported in Table~\ref{tab:nexp}.
First, we are bound to consider the case
that MACHO lensing does not contribute at all,
so that we are left with the expected lensing signal
from the luminous populations alone, with
an expected number of events for self-lensing 
somewhat smaller than 1. Even allowing for both
our candidates to be genuine microlensing,
this is still, according to  Poisson statistic, 
fully compatible with the observations.

On the other hand, the expected number of MACHO lensing events is not large
if compared to that of self lensing. Even for a \emph{full} 
MACHO halo we would expect about as many self-lensing events
as MACHO lensing. It is clear, therefore, that on the basis
of the number of events alone our statistics
are still too small to draw conclusions on this contribution.

It may be asked, then, how we might disentangle the two signals:
self lensing from MACHO lensing.
First, an improvement in the event statistics
might help us to further exploit the expected
differences in the spatial distributions
(top panels, Fig.~\ref{fig:mls}).
Second, a good enough sampling 
(such as was not the case for the present selection)
might allow a sufficient
characterisation of the selected flux variations. In turn,
this could give possible hints on the nature of the lens,
as was the case for the detailed analysis linked
to the finite  source size effect for the PA-S3 event
carried out by \cite{arno08}.
Both these reasons motivate the need
for further observations.

\section{Conclusions} \label{sec:end}

In this paper we have presented a thorough analysis
of the data acquired during the
second season of our pixel-lensing observational campaign
carried out towards M31 with the $152~\mathrm{cm}$ Cassini telescope
located in Loiano. We have developed a fully automated pipeline analysis
for the detection and the characterisation of
\pacz\ like microlensing flux variations. We have 
evaluated, through a Monte Carlo simulation,
the expected signal, given an astrophysical
model for M31 and the efficiency of our pipeline.
As a result we have selected 1-2 candidate microlensing events.
This output turns out to be in fair agreement
with our evaluation of the expected self-lensing signal from
the luminous M31 components. 
As for the would-be MACHO lensing signal,
the statistics of events are still too small
to draw firm conclusions. 
Further observations might help us to disentangle
between the two populations. First, because
of the larger statistics
of events, second (provided a better sampling) 
because of a better characterisation of the light curves.

Previous analyses, in particular those
carried out using data collected at the INT telescope
by the POINT-AGAPE and the MEGA collaborations,
have reached different conclusions on the dark matter
halo content in the form of MACHOs along the line of sight
towards M31 \citep{novati05,mega06}. Their fundamental point of disagreement
may be traced back to the evaluation of the
expected self-lensing signal. In this perspective
our campaign, as well as that carried out by 
the ANGSTROM collaboration right towards the M31 center
\citep{kerins06}, may help in shedding more light onto
this important issue.

\acknowledgments
We thank the anonymous referee for useful comments
and suggestions.
The observational campaign has been possible
thanks to the generous allocation of telescope
time by the TAC of the Bologna Observatory
and to the invaluable help of the technical staff.
We thank L.~Inno, A.~Klein, M.~Miranda, R.~Ungaro and
L.~Santoro  for taking part to the observations.
SCN acknowledges support for this work by the Italian Space Agency (ASI).
SCN, VB, LM and GS acknowledge support by MIUR
through PRIN 2006 prot. 2006023491\_003, and by
FARB of the Salerno University. 
MS is supported by the Swiss National
Science Foundation and the dr. Tomalla Foundation.
This work has made use of the IAC-STAR synthetic color-magnitude
diagram computation code. IAC-STAR is supported and maintained
by the computer division of the Instituto de Astrofisica de Canarias.
We thank the POINT-AGAPE collaboration for access to their database.

\appendix

\section{The microlensing pipeline: the sampling criterion} \label{sec:app_t0}

The rationale of the present cut is to 
test whether the detected flux variation
is sufficiently sampled. As a starting point
we use the results of the \pacz\ fit, and
in particular the values of the time at maximum
magnification, $t_0$, and of the bump duration,
the full-width-at-half-maximum $t_{1/2}$.
Given these values we identify six intervals
along the light curve, symmetric around $t_0$,
$[t<t_0-3~t_{1/2}]$, 
$[t_0-3~t_{1/2}<t<t_0- t_{1/2}/2]$, 
$[t_0-t_{1/2}/2<t<t_0]$, 
$[t_0<t<t_0+t_{1/2}/2]$,
$[t_0+t_{1/2}/2<t<t_0+3~t_{1/2}]$ and
$[t>t_0+3~t_{1/2}]$.

As a first set of criteria we ask for at least $n_\mathrm{min}$
data-points in at least 3 out of the 4 inner intervals,
within $t_0\pm 3~t_{1/2}$,
\emph{and} at least 1 data-point in \emph{both} the joined
outer intervals $t<t_0-t_{1/2}/2$ and
$t>t_0+t_{1/2}/2$, namely, we allow the same data-point to be counted twice.
The value of $n_\mathrm{min}$ is fixed
according to the duration, with $n_\mathrm{min}=1,2,3$
for $t_{1/2}<5, <15, >15~\mathrm{days}$, respectively.

As a second set of criteria we ask for at least 1 data-point
in at least 2 out of the 4 inner intervals
\emph{and} at least 1 data-point in at least 1 of
the two tail intervals ($t<t_0-3~t_{1/2}$ and
$t>t_0+3~t_{1/2}$). 

We refer to the two set of criteria,
and to the corresponding selected flux variations,
as set ``A'' and set ``B'', respectively.

The set A criterion is more demanding 
as for the sampling along the inner part of the flux variation,
whereas set B makes, in a way, a stronger demand
on the coverage of the far tails, so that
set A selected events are not a simple subsample
of set B events. Furthermore, it results that
the time of maximum magnification for set A events
is constrained within the limits of the observational run
whereas for set B it can also fall (slightly) outside.
Set B flux variations therefore enjoy  an effective longer baseline.

The sampling analysis is carried out along $R$-band data only.

\section{The microlensing pipeline: the PSF criterion} \label{sec:app_psf}

As outlined in Sect.~\ref{sec:anaml},
through this criterion we want to check
whether the detected variation
along the light curve can be attributed
to a physically meaningful flux variation
(a variable star or microlensing)
or to  some kind of spurious signal
(cosmic rays, bad pixel, seeing effects and so on).
To this purpose we have to turn to an analysis
carried out on the images. As a criterion we test whether the \emph{spatial} form
of the detected bump has a well enough shaped PSF. As we are not carrying out
difference image photometry, we can safely stop 
at the first order approximation
of a 2-dimensional Gaussian, namely
\begin{equation} \label{eq:psf}
P(x,y)= \exp
\Bigg\{
-\frac{1}{2(1-\rho^2)} 
\Bigg[
\left(\frac{x-\mu_x}{\sigma_x}\right)^2+
\left(\frac{y-\mu_y}{\sigma_y}\right)^2-
2\rho
\left(\frac{x-\mu_x}{\sigma_x}\right)
\left(\frac{y-\mu_y}{\sigma_y}\right)
\Bigg]
\Bigg\}\,.
\end{equation}

We carry out this analysis on a difference image  
(a window of 19x19 pixels around the central pixel $x_0,y_0$
in which the flux variation has been detected)
built as follows. Given the time of maximum magnification
of the bump variation, $t_0$, we look for the nearest night
of observation, and we average the images collected during
this night, taking note of the average seeing of the night.
As a baseline image we take the average of a set of images chosen 
along the baseline with a similar seeing.

As a criterion, we ask for the fit 
with respect to the 2-dimensional Gaussian 
to converge ``properly''\footnote{We make use of
the HFITH function of the CERNLIB.}. Besides the actual demand
for the fit to converge we also ask
the following : $|x_0-\mu_x|<1.5$, $|y_0-\mu_y|<1.5$,
$1<\sigma_x<3$, $1<\sigma_y<3$ and $|\rho|<0.5$.
This additional set of criteria is necessary as it is very
easy, in case of spurious variations, for the fit to converge
while it slips at the border of 
the physically meaningful bounds.

It is worth noticing that a different
parametrisation for the 2-d Gaussian profile
with respect to that chosen in Eq.~\ref{eq:psf}
is sometime used, namely
\begin{equation} \label{eq:psf2}
P_1(x,y)= \exp\Bigg[-\frac{(x\cos\phi+y\sin\phi)^2}{2 \sigma_a^2}
-\frac{(x\sin\phi-y\cos\phi)^2}{2 \sigma_b^2}\Bigg]\,.
\end{equation}
In fact the two formulations are equivalent, one can easily
evaluate the parameters $\sigma_a,\,\sigma_b$ and $\phi$ 
as a function of $\sigma_x,\,\sigma_y$ and $\rho$, and vice versa
(it must be noted, however, that the parameter $\rho$
plays a double role, giving both an inclination and a distortion).
Within the parametrisation of Eq.~\ref{eq:psf2} we find
our cuts to roughly correspond to $1<\sigma_a<3$, $1<\sigma_b<3$, 
and $\phi\approx\mathrm{arbitrary}$.  Still, we prefer Eq.~\ref{eq:psf} 
as the fit procedure looks, in this case, more stable.

\section{Monte Carlo simulation: the drawing process} \label{sec:app_mls}

In this Appendix we discuss some further details
regarding the Monte Carlo simulation described in Sect.~\ref{sec:mls}.

A microlensing event is fully specified by 10 parameters.
The line of sight, specified by two angles, $\theta,\rho$,
out of which we determine the (angular) position
within our fields as $x=\rho \cos(\theta), y=\rho \sin(\theta)$;
the lens distance $D_l$ and mass $\mu_l$;
the source distance $D_s$ and flux $\phi_s$; the lens-source
relative velocity, specified by a modulus $v_r$ and a phase $\psi_v$;
finally the microlensing amplification parameters,
the impact parameter $u_0$ and the time of maximum amplification $t_0$.
Each of these parameters has its own physical distribution function
from which we draw a value for each event realisation.
Even if we simulate only $R$ light curves, 
the available information within the luminosity functions used
allows us to evaluate also the source radius,
so to properly take into account the finite size source effect, 
as for the microlensing amplification.

A key aspect of our Monte Carlo scheme is the ``weight'', $w_i$, that
we associate to each event. 
This is linked in part to the drawing process and in part
to the physical process we are studying.
As for the first part, this follows from the fact that
for a given distribution function we have two choices as for
the drawing process. Either we may directly draw according
to the distribution, and in this case the corresponding weight
is $w_i=1$, either we may draw with a uniform distribution
and give the event a weight proportional to the distribution function.
As for the second part, we take the weight
to be proportional to the following
\begin{equation} \label{eq:weight}
w_i\propto \left(2\,R_\mathrm{E}\,u_\mathrm{MAX}\,v_r\,\Delta T_\mathrm{OBS}\right) \times n_s\,,
\end{equation}
where $R_\mathrm{E}$ is the Einstein radius, $u_\mathrm{MAX}$ the maximum value
for the impact parameter, $\Delta T_\mathrm{OBS}$ the duration of the observational campaign,
$n_s$ the number surface density of available sources (fox a fixed line of sight).
In fact, according to our simulation scheme,
first we fix the \emph{lens} line of sight and specify all of the events
characteritics. Then , as the purpose of the Monte Carlo is to evaluate the \emph{number}
of expected microlensing events, we have to evaluate
how many sources (given the M31 surface brightness and luminosity function)
are available within a surface delimited by the event characteristics 
and the experimental conditions. Indeed, the term
$2\,R_\mathrm{E}\,u_\mathrm{MAX}\,v_r\,\Delta T_\mathrm{OBS}$ in Eq.~\ref{eq:weight}
is the surface area on the lens plane (for a fixed event configuration) 
where it is possible to find a suitable source. 

The last contribution to the weight comes from the need to evaluate
the number of available lenses. For each lens population
this is computed as the ratio of the total mass of the component
within the observed field of view divided by the average lens mass.
More specifically, the weight linked to the choice
of the line of sight is normalized to the total mass.

As for the drawing of the line of sight we have the following:
for lenses in the Milky Way halo we assume the spatial distribution
across the observed field of view to be uniform;
for lenses in the M31 halo, because of the assumed spherical symmetry,
the resulting distribution for the angle $\theta$ is uniform;
for bulge and disk lenses we make use of the 
\cite{kent89} bulge-disk decomposition
(whenever using the \cite{kent89} profile we use
a uniform distribution and then attribute to the
event a ``weight'' as described above).

Once the event configuration, and its weight, have been specified, we build
the corresponding light curve and, given our observational set up,
we evaluate whether the event is selected or not.
This selection process is completely independent of the event weight.
The selected events delimit the overall integration space,
so that the expected number of events is given by the sum
of the weights of these selected events only. (This same set
of events is then taken, with all of their characteristics,
as the input set for the simulation on the images
described in Sect.~\ref{sec:efficiency}.)

In Table~\ref{tab:mls} we report, for each parameter,
the distribution function used and its range of variability.

\begin{deluxetable}{rrrrr|r}
  \tablewidth{15cm}
  \centering
\tablecolumns{6}
\tablewidth{0pc}
\tablecaption{Monte Carlo parameters:
distributions function and limits for each lens/source population.\label{tab:mls}}
\tablehead{
 & \colhead{MW halo} & \colhead{M31 halo} & \colhead{M31 bulge} & \colhead{M31 disk}&\\
 &&&&&limit\\
}
\startdata
$\theta$ & uniform & uniform & K89& K89+$f(\zeta)$& field of view\\
$\rho$ &uniform & $\rho_\mathrm{HM31}(\vec{r})$ & K89&K89+$f(\zeta)$&field of view\\
$D_l$ &$\rho_\mathrm{HMW}(\vec{r})$ & $\rho_\mathrm{HM31}(\vec{r})$ &K89&K89+$f(\zeta)$& model\\
$\mu_l$ & $\delta(\mu_l)$ & $\delta(\mu_l)$ & $\mu^{-\alpha}$ & $\mu^{-\alpha}$& model\\
$D_s$ & -&-& K89 &K89+$f(\zeta)$& model\\
$\phi_s$ &-&-& IAC & Hipparcos& $M_I<2$\\
$v_r$ &Gaussian&Gaussian&Gaussian&Gaussian& $0,\infty$\\
$\psi_v$ &uniform&uniform&uniform&uniform& $0,2\pi$\\
$u_0$ &uniform&uniform&uniform&uniform& $0,u_\mathrm{MAX}(=2)$\\
$t_0$ &uniform&uniform&uniform&uniform& $0,\Delta T_\mathrm{obs}$\\
\enddata
\tablecomments{
The parameters have been introduced in Appendix~\ref{sec:app_mls}.
The details for the density function
of the Milky Way and the M31 halos and for the source luminosity functions
are given in Sect.~\ref{sec:model}.
K89 stands for the \cite{kent89} bulge-disk decomposition.
For the disk, as only a two-dimensional distribution
on the plane is given, we consider a vertical distribution
$f(\zeta) = \mathrm{cosh}(-(\zeta/\zeta_h))^2$, where $\zeta_h$ is the
vertical scale height.
The index for the power law mass function are given in  Sect.~\ref{sec:model}.
In the last column we report the range of variability for each parameter:
for $\theta,\rho$ ``field of view'' indicates the two observed fields, ``Nord'' and ``Sud''
(Sect.~\ref{sec:setup});
for the distances we take the physical boundaries of the component considered,
in particular for the halos we use a truncation radius as specified in Sect.~\ref{sec:model},
for the bulge $\sim~7~\mathrm{kpc}$ around the M31 distance while for the disk vertical component
we draw directly from the given distribution (for a fixed line of sight
this allows us to evaluate the physical distances);
for bulge and disk lenses $\mu_l$ is drawn with a lower limit
of $0.08~\mathrm{M}_\odot$ and upper limit of $1.0~\mathrm{M}_\odot$ and
$10.0~\mathrm{M}_\odot$, respectively.
}
\end{deluxetable}

\bibliographystyle{apj}
\bibliography{biblio}

\end{document}